\documentclass[a4paper,fleqn,oneside,12pt]{article}
\usepackage{amsmath}
\usepackage{amssymb}
\usepackage{amsfonts}
\usepackage{amsthm}
\usepackage{mathrsfs}
\usepackage{latexsym}
\usepackage{graphicx}
\usepackage[german,greek,english]{babel}
\usepackage[T1]{fontenc}
\newcommand{\tl}{\textlatin}
\newcommand{\be}{\begin{eqnarray}}
\newcommand{\ee}{\end{eqnarray}}
\newcommand{\bse}{\begin{subequations}}
\newcommand{\ese}{\end{subequations}}
\newcommand{\bdm}{\begin{displaymath}}
\newcommand{\edm}{\end{displaymath}}
\theoremstyle{definition}

\newtheorem*{criterion}{Criterion}

\numberwithin{equation}{section}
\begin{document}
\title{\Large\textbf{On the essential constants in Riemannian geometries}}
\author{\textbf{G.O. Papadopoulos}\thanks{E-Mail: gopapado@phys.uoa.gr}}
\date{}
\maketitle
\begin{center}
\textit{National \& Kapodistrian University of Athens, Physics Department\\
Nuclear \& Particle Physics Section\\
Panepistimioupolis, Ilisia GR 157--71, Athens, Hellas}
\end{center}
\begin{center}
\textit{\textbf{Dedication}}
\textit{The author dedicates this work to Elena and Katerina}.
\end{center}
\begin{abstract}
In the present work the problem of distinguishing between essential and spurious (i.e.,
absorbable) constants contained in a metric tensor field in a Riemannian geometry is considered.
The contribution of the study is the presentation of a sufficient and necessary criterion, in
terms of a covariant statement, which enables one to determine whether a constant is essential or
not. It turns out that the problem of characterization is reduced to that of solving a system of
partial differential equations of the first order. In any case, the metric tensor field is
assumed to be smooth with respect to the constant to be tested.\\
It should be stressed that the entire analysis is purely of local character.

\vspace{0.3cm}
\noindent
\textbf{MSC-Class} (2000): 83C05;83C15;83C20
\end{abstract}
\section{Introduction}
When dealing with Riemannian spaces, especially in a local description and in a coordinate
approach, one frequently encounters the problem of attributing a character to a constant (or a
parameter) which may appear in the metric tensor field. Generally, there are two possibilities:
this constant is either essential (i.e., a true degree of freedom) or spurious (i.e., absorbable
with the help of a change in the coordinates).\\
The issue is of great interest in the context of general relativity, where the metric tensor
field is the solution to the Einstein equations, and the constants emerge from the integration
procedure. But, this observation will not limit the spirit of the present work.

There is a variety of ways to attack the problem under discussion. In the relevant literature,
one can find two main approaches:
\begin{itemize}
\item[$A_{1}$] The first main approach consists simply in trying to find that particular change
in the coordinates which can serve to eliminate the ``suspect'' constant. When this is possible,
the constant is incorporated in the very definition of the new coordinate system, being thus
absorbed. The difficulty here is that, in general, there is no systematic way to find the desired
transformation. Obviously, failure to find such a transformation does not necessarily imply
the essentiality of the constant.
\item[$A_{2}$] The second main approach, which is more elaborate and sophisticated, can be
divided into two subcategories: one can either use the invariant classification methods for a
single Riemannian space, or implement the methods of the \textit{equivalence problem}
(ref.\ \cite{Stephani_et_al.}).
The second way (which may be more laborious than the first) consists in the following steps: one
considers twice the metric tensor field: once for a given value of the constant and once for
another value of it. The final step is to compare these two metrics and to check whether they are
equivalent or not. A positive answer dictates that the constant is spurious (and a negative, that
it is essential) (see also ref.\ \cite{Geroch} for a connection between limits of space-time and
the problem of essentiality).
\end{itemize}

The non-equivalence between two given Riemannian spaces can easily be checked using the notion of
\textit{curvature invariant relations}, functionally independent relations among scalars. These
scalars are constructed either from the Riemann tensor and its covariant derivatives up to a
given order by contracting all the indices, or as ``ratios'' between two tensors (obtained from
the Riemann tensor) which differ only by a factor. The first case gives scalars entering
\textit{syzygies}, \textit{polynomial invariants}, \textit{mixed invariants}, and the
\textit{Cartan invariants} --see ref.\ \cite{Stephani_et_al.} and the references therein for
details. The second case is described in ref.\ \cite{Kundt} (especially the last two references
therein).
It is sufficient for the two given spaces to differ in only one such relation in order to be
inequivalent.\\
Curvature-invariant relations have one very important property: they do not depend on points of
the Riemannian space; thus, their functional forms are invariant statements (i.e., they retain
the same functional form in all coordinate systems). Consequently, if these functional forms
depend on some constants, that means these constants, which clearly are some from those in the
metric tensor field, are not affected by a change in the coordinates. If a constant of the
metric tensor field could be eliminated by such a change, then the metric tensor field in the new
coordinate system as well as all the curvature invariant relations based on it, would lack this
particular constant, an invariant statement, since the curvature invariant relations are
invariant in form. Therefore, only essential constants will appear in the curvature-invariant
relations.\\
An example will elucidate the above arguments: consider the well-known Schwarzschild metric in
the usual local coordinate system $\{t,r,\theta,\phi\}$ and e.g., the two curvature scalars:
\be
S_{1}&\equiv & R_{\alpha\beta\mu\nu}R^{\alpha\beta\mu\nu}=48M^{2}/r^{6}\\
S_{2}&\equiv & S^{~;\mu}_{1;\mu}=-3456M^{3}/r^{9}+1440M^{2}/r^{8}
\ee
which are, of course, $r$-dependent. However, if $r$ is eliminated between them, one arrives at
the relation:
\be
R(S_{1},S_{2},M)\equiv S_{2}+6\sqrt{3}~S_{1}^{3/2}-5M^{-2/3}\sqrt[3]{9/2}~S_{1}^{4/3}=0
\ee
This relation is not only independent of the space-time points, i.e., it can be evaluated
everywhere in the Schwarzschild space-time (except, of course, the true singularity at $r=0$),
but also invariant under any change in the local coordinate system; although the functional form
(in terms of the coordinates) of the two curvature scalars $S_{1}$, $S_{2}$ will change, the
relation $R(S_{1},S_{2},M)$ will retain its form (as a function of its arguments $S_{1}$,
$S_{2}$, and $M$) and thus constitutes a curvature invariant relation. Indeed, consider for
example the change $r\rightarrow \widetilde{r}: r=M^{1/3}\widetilde{r}$ which eliminates the
parameter $M$ from $S_{1}$ and alters the form of $S_{2}$, yet keeps the relation
$R(S_{1},S_{2},M)$ unchanged.\\
To use the above considerations in order to deduce equivalence between two Riemannian spaces is
problematic, since it would require the existence of a countable basis for an arbitrary
functional space.

The following section presents a sufficient and necessary criterion, in a covariant language
which offers one the ability to check whether a constant, appearing in a metric tensor field, is
essential or not. In the second case, the criterion also provides a way to find the desired local
finite transformation of the change in the coordinates.

\section{The Criterion}
Before presenting the criterion, a word must be said for the existence of yet another kind of
constant, namely the global (or topological) constants: indeed, there are cases where a constant
can be removed from local coordinate patches but it does appear in the transforms between them
(e.g., in the appropriate range of the coordinates).

From the previous section, it is clear that \textit{essential} and \textit{spurious} are mutually
complementary notions. It will turn out more practical, though equivalent, to deal with
spurious. Indeed, if the constant is spurious one can, in the coordinates in which the constant
is removed, take a product metric tensor field on $\mathcal{S}\times I$ (where $\mathcal{S}$ is
the initial $n$-dimensional manifold and $I$ the domain of definition of the spurious constant),
and then deduce that the only non-zero components of curvature in $n+1$ dimensions are those
which correspond to the curvature tensor of the $n$-dimensional metric tensor field. In these
coordinates the normals to $\mathcal{S}$ form a symmetry. Alternatively, one can also consider
the $(n+1)$-dimensional manifold using the original coordinates, with the spurious constant as
the extra coordinate, and use the constant to label the $n$-dimensional slices.\\
The above arguments can be made more precise as follows:\\
Let $\mathcal{S}$ be a Riemannian space which is described by the pair $(\mathcal{M},g)$,
where\footnote{This definition is influenced by the definition of space-time, but $C^{\infty}$
--instead of simply $C^{r}$, connectedness as well as the Hausdorff condition seem to be minimal
extensions.} $\mathcal{M}$ is an $n$-dimensional, connected, Hausdorff and $(C^{\infty})$
manifold and $g$ is a ($C^{r}$)\footnote{The value of $r$ depends on the applications. In the
context e.g., of general relativity, it is assumed that $r\geq 2$ --see ref.\
\cite{Hawking and Ellis}, pp.\ 55-59 for a relevant discussion.} metric tensor field on it; a
non-degenerate, covariant tensor field of order 2, with the property that at each point of
$\mathcal{M}$ one can choose a frame of $n$ vectors $\{z_{0},\ldots,z_{n-1}\}$, such
that\footnote{Small Greek indices take the values \{0,\ldots,$n-1$\}.}:
$g(z_{\alpha},z_{\beta})=\eta_{\alpha\beta}$, where $\eta$ is a diagonal matrix with entries
$\{\varepsilon_{0},\ldots,\varepsilon_{n-1}\}$, and: $\varepsilon_{\alpha}=\pm 1$.\\
Let also this metric tensor field depend on a constant $\lambda$; so in a local coordinate
system $\{x^{\mu}\}$, it is:
\be
g_{\alpha\beta}=g_{\alpha\beta}(x^{\gamma};\lambda)
\ee
It is also supposed that the metric tensor field $g$ is a ($C^{\infty}$) function (i.e., smooth)
with respect to $\lambda$ --a basic assumption which is also encountered in ref.\ \cite{Geroch},
where limits of space-time are considered (which of course have to be defined in terms of
essential constant(s)).\\
Let $I\subseteq \mathbb{R}$ be the domain of definition (i.e., the range of possible values) of
the constant $\lambda$. Another Riemannian space $\mathcal{\widetilde{S}}$ can, naturally,
emerge; the product: $\mathcal{\widetilde{S}}=\mathcal{S}\times I$. By this it is meant that the
initial Riemannian space is nothing but the hypersurface $\lambda=\text{const.}$ in
$\mathcal{\widetilde{S}}$; a local isometric embedding.\\
If $p \in \mathcal{\widetilde{S}}$, then the tangent space $T_{p}\mathcal{S}$ of $\mathcal{S}$
is a subspace of $T_{p}\mathcal{\widetilde{S}}$. Since $\mathcal{S}$ is a regular submanifold of
$\mathcal{\widetilde{S}}$, there exists a basis\footnote{Capital Latin indices take the values
\{0,\ldots,$n$\}.} $\{\mathbf{e}_{0},\mathbf{e}_{\mu}\}\equiv \mathbf{e}_{N}$ of
$T_{p}\mathcal{\widetilde{S}}$  such that its ``spatial'' part $\{\mathbf{e}_{\mu}\}$ is the
basis of $T_{p}\mathcal{S}$. Since the difference of dimensions is 1, the subspace has no
torsion. Consequently there is only one normal to it, vector $\mathbf{n}$. Without loss of
generality it is taken to be of unit length. Then, one assigns:
\be
\mathbf{n}=n^{A}\mathbf{e}_{A}=\frac{1}{N}\mathbf{e}_{0}
-\frac{N^{\alpha}}{N}\mathbf{e}_{\alpha} \Rightarrow
n^{A}=\frac{1}{N}\left\{1,-N^{\alpha}\right\}
\ee
with:
\be
\widetilde{g}(\mathbf{n},\mathbf{n})=\varepsilon=\pm 1
\ee
(the sign is rather irrelevant), so:
\be
\mathbf{e}_{0}=N\mathbf{n}+N^{\alpha}\mathbf{e}_{\alpha}
\ee
The quantity $N$ is the \textit{lapse function} and the object $N^{\alpha}$ is the
\textit{shift vector}.
By definition:
\begin{align}
\widetilde{g}_{00}&=\widetilde{g}(\mathbf{e}_{0},\mathbf{e}_{0})
=N^{2}\widetilde{g}(\mathbf{n},\mathbf{n})
+N^{\alpha}N^{\beta}\widetilde{g}(\mathbf{e}_{\alpha},\mathbf{e}_{\beta})\\
\widetilde{g}_{0\alpha}&=\widetilde{g}(\mathbf{e}_{0},\mathbf{e}_{\alpha})
=N_{\alpha}\equiv N^{\beta}g_{\alpha\beta}\\
\widetilde{g}_{\alpha\beta}&=\widetilde{g}(\mathbf{e}_{\alpha},\mathbf{e}_{\beta})
=g_{\alpha\beta}
\end{align}
Thus, the Greek indices change position with the initial metric $g_{\alpha\beta}$, while the
capital Latin indices change position with the new metric $\widetilde{g}_{AB}$.\\
Finally:
\be
\widetilde{g}_{AB}=\begin{pmatrix}
  N_{\rho}N^{\rho}+\varepsilon N^{2} & N_{\beta} \\
  N_{\alpha} & g_{\alpha\beta}
\end{pmatrix}
\ee
A straightforward calculation results in:
\bse
\begin{align}
\widetilde{\Gamma}^{0}_{00}&=\frac{\dot{N}}{N}+\varepsilon\frac{N^{\mu}N^{\nu}}{N}K_{\mu\nu}
+\frac{N^{\mu}}{N}N_{|\mu}\\
\widetilde{\Gamma}^{0}_{0\nu}&=\varepsilon\frac{N^{\mu}}{N}K_{\mu\nu}+\frac{N_{|\nu}}{N}\\
\widetilde{\Gamma}^{0}_{\mu\nu}&=\varepsilon\frac{K_{\mu\nu}}{N}\\
\widetilde{\Gamma}^{\kappa}_{00}&=-\frac{\dot{N}}{N}N^{\kappa}
-\varepsilon\frac{N^{\mu}N^{\nu}N^{\kappa}}{N}K_{\mu\nu}
-\frac{N^{\mu}N_{|\mu}}{N}N^\kappa+\dot{N}^\kappa\notag\\
&-\varepsilon NN^{|\kappa}+N^{\kappa}_{|\nu}N^{\nu}-2NK^{\kappa}_{\nu}N^{\nu}\\
\widetilde{\Gamma}^{\kappa}_{0\nu}&=-\frac{N^{\kappa}}{N}N_{|\nu}
-\varepsilon\frac{N^{\kappa}N^{\mu}}{N}K_{\mu\nu}+N^{\kappa}_{|\nu}
-NK^{\kappa}_{\nu}\\
\widetilde{\Gamma}^{\kappa}_{\mu\nu}&=\Gamma^{\kappa}_{\mu\nu}
-\varepsilon\frac{N^{\kappa}}{N}K_{\mu\nu}
\end{align}
\ese
where:
\be
K_{\mu\nu}=\frac{1}{2N}\left(N_{\mu|\nu}+N_{\nu|\mu}-\dot{g}_{\mu\nu}\right)
\ee
is the \textit{extrinsic curvature} (in the literature of the theory of surfaces, it is also
known as \textit{second fundamental form}, of \textit{shape tensor}) and describes the
embedding curvature.\\
The bar (\tl{|}) denotes covariant derivative with respect to the initial metric $g$ of the
subspace, while the dot ($\cdot$) denotes differentiation with respect to the extra coordinate,
i.e., $\lambda$.\\
The general theory of embedding can be found in any book on differential geometry, e.g.,
\cite{Eisenhart_O'Neil} are some classical references. There, one can see that the present case,
where the difference in the dimensions is 1 (resulting in zero torsion for the subspace) is very
simple. In fact, the Mainardi-Codazzi conditions are identically satisfied, while the
Weingarten-Gauss conditions assume the form:
\bse\label{Weingarten-Gauss}
\begin{align}
\widetilde{R}_{\alpha\beta\mu\nu}&=R_{\alpha\beta\mu\nu}-\varepsilon(K_{\alpha\mu}K_{\beta\nu}
-K_{\alpha\nu}K_{\beta\mu})\\
\widetilde{R}_{\perp \beta\mu\nu}&=K_{\beta\nu|\mu}-K_{\beta\mu|\nu}
\end{align}
\ese
of course, after the use of the projections:
\be
T^{A\ldots}_{B\ldots}n^{B}\equiv T^{A\ldots}_{\perp\ldots},\quad
T^{A\ldots}_{B\ldots}n_{A}\equiv T^{\perp\ldots}_{B\ldots},\quad
T^{A\ldots}_{B\ldots}y^{B}_{,\alpha}\equiv T^{A\ldots}_{\alpha\ldots}
\ee
$y^{B}_{,\alpha}$ being the Jacobian $\partial y^{A}/\partial x^{\alpha}$ between a set of local
coordinates in $\mathcal{\widetilde{S}}$, say $\{y^{A}\}$, and the set of the corresponding local
coordinates in $\mathcal{S}$, say $\{x^{\alpha}\}$.\\
For the chosen embedding it is: $\{y^{A}\}=\{\lambda,x^{\alpha}\}$

If one defines the tensor on $\mathcal{\widetilde{S}}$:
\be
C_{AB}\doteq-\frac{1}{2}\pounds_{\mathbf{n}}\widetilde{g}_{AB}\equiv
-\frac{1}{2}(\mathbf{n}_{A;B}+\mathbf{n}_{B;A})
\ee
where the semicolon (;) denotes covariant differentiation with respect to the new metric
$\widetilde{g}$, one will have:
\be
C_{AB}=\begin{pmatrix}
 N^{\mu}N^{\nu}K_{\mu\nu}+\varepsilon N^{\mu}N_{|\mu} &
 \varepsilon\frac{1}{2}N_{|\beta}+K_{\beta\mu}N^{\mu} \\
 \varepsilon\frac{1}{2}N_{|\alpha}+K_{\alpha\mu}N^{\mu}  & K_{\alpha\beta}
\end{pmatrix}
\ee
In order for the two spaces, i.e., the embedding and the embedded, to have exactly the same
geometrical information (in other words, exactly the same curvature properties), something which
happens when and only when the constant (i.e., the extra coordinate) $\lambda$ is absorbable, the
Weingarten-Gauss conditions \eqref{Weingarten-Gauss} suggest that the extrinsic curvature must
vanish --for any embedding:
\be\label{Demand}
K_{\alpha\beta}=0
\ee
Condition \eqref{Demand} as well as the demand for its validity for any embedding, and thus for
the particular embedding in a Gaussian system of coordinates: $N=1$ or $N=N(\lambda)$ and
$N^{\alpha}=0$, result in the vanishing of the tensor $C_{AB}$; an invariant statement. Hence,
follows the:
\begin{criterion}
The constant $\lambda$ contained in the metric tensor field $g$ of the Riemannian space
$\mathcal{S}$ is spurious, if and only if the Lie derivative of the metric tensor field
$\widetilde{g}$ of the embedding space $\mathcal{\widetilde{S}}$ with respect to the normal
(to the subspace) vector $\mathbf{n}$, $\pounds_{\mathbf{n}}\widetilde{g}$, vanishes.
\end{criterion}
\begin{proof}
First, one observes that the vanishing of the tensor field $C_{AB}$ results in the following
set of partial differential equations (PDEs):
\bse
\begin{align}
C_{00}=0 &\Rightarrow N^{\mu}N_{|\mu}=0 \\
C_{0\alpha}=0 &\Rightarrow N_{|\alpha}=0\\
C_{\alpha\beta}=0 & \Rightarrow K_{\alpha\beta}=0
\end{align}
\ese
or:
\bse
\begin{align}
&N=N(\lambda) ~~~\text{(though an arbitrary function)}\\
&N_{\alpha|\beta}+N_{\beta|\alpha}=\dot{g}_{\alpha\beta}\label{Equation}
\end{align}
\ese
The lines preceding the criterion prove its necessity. In order to prove its sufficiency, let
$n^{A}=\frac{1}{N(\lambda)}\{1,-N^{\alpha}(\lambda, x^{\beta})\}$ a normal vector whose
components satisfy \eqref{Equation}.
The set of its integral curves, parametrized by a parameter $s$, has the form:
\be
\frac{dy^{A}}{d s}=n^{A}(y^{B}(s))
\ee
and, from the theory of ordinary differential equations, it is known that this problem is well
posed and it always has a solution. Written out in detail:
\bse\label{Flows}
\begin{align}
&\frac{dy^{0}}{d s}=\frac{1}{N(y^{0})}\\
&\frac{dy^{\alpha}}{d s}=-\frac{N^{\alpha}(y^{0},y^{\beta})}{N(y^{0})}
\end{align}
\ese
As usual, this set defines a one-parametric ($s$ being the parameter) family of transformations
from the set $\{y^{A}\}$ to the set $\{\overline{y}^{A}\}$, the latter being the constants of
integration of the flow lines of the vector $\mathbf{n}$. It is very easy to see that the
emerging transformation has the general functional form:
\bse\label{Form_of_the_Transformation}
\begin{align}
&y^{0}\rightarrow \overline{y}^{0}: \overline{y}^{0}=f(y^{0})\\
&y^{\alpha}\rightarrow \overline{y}^{\alpha}: \overline{y}^{\alpha}=f^{\alpha}(y^{0},y^{\gamma})
\end{align}
\ese
while the vector $\mathbf{n}$ undergoes a change:
\be
\mathbf{n}\rightarrow \overline{\mathbf{n}}:
\overline{n}^{A}=\frac{\partial \overline{y}^{A}}{\partial y^{B}}n^{B}
\ee
with the help of the transformation \eqref{Form_of_the_Transformation}:
\be
\overline{n}^{A}=\frac{1}{N(y^{0})}\Big\{\frac{\partial f(y^{0})}{\partial y^{0}},
\frac{\partial f^{\alpha}(y^{0},y^{\gamma})}{\partial y^{0}}
-\frac{\partial f^{\alpha}(y^{0},y^{\gamma})}{\partial y^{\beta}}N^{\beta}(y^{E})\Big\}
\ee
But:
\be
\frac{d \overline{y}^{\alpha}}{d s}=0\Rightarrow
\frac{\partial f^{\alpha}(y^{0},y^{\gamma})}{\partial y^{0}}
-\frac{\partial f^{\alpha}(y^{0},y^{\gamma})}{\partial y^{\beta}}N^{\beta}(y^{E})=0
\ee
by virtue of the flow lines equations \eqref{Flows}. Thus:
\be
\mathbf{n}\rightarrow \overline{\mathbf{n}}:
\overline{n}^{A}=\frac{1}{N(y^{0})}\Big\{\frac{\partial f(y^{0})}{\partial y^{0}},0\Big\}
\ee
i.e., a Gaussian system of coordinates. Hence, in the new coordinate system, the vanishing of the
tensor $\overline{C}_{AB}$, obviously, is tantamount to:
\be
\frac{\partial \overline{g}_{\alpha\beta}}{\partial \overline{y}^{0}}=0
\ee
i.e., the transformed metric tensor field of the subspace does not contain the corresponding
extra coordinate $\overline{y}^{0}$, which is a function of the constant under discussion.
\end{proof}

\section{An application and a pedagogical example}
One immediate and simple application of the criterion is achieved when the latter is applied to
the case where the ``suspect'' constant is an overall factor; i.e., in a local system of
coordinates $\{x^{\mu}\}$:
\be
g_{\alpha\beta}=g_{\alpha\beta}(x^{\gamma};\lambda)\equiv \lambda G_{\alpha\beta}(x^{\gamma})
\ee
Then, the criterion, in its ``solved form'' \eqref{Equation}, results in:
\be
N_{\alpha|\beta}+N_{\beta|\alpha}=\dot{g}_{\alpha\beta}=G_{\alpha\beta}\Rightarrow
N_{\alpha|\beta}+N_{\beta|\alpha}=\frac{1}{\lambda}g_{\alpha\beta}
\ee
which is nothing but the homothety equations for the subspace --a well-known result.

For the sake of simplicity and brevity, the paper concludes with a pedagogical example.\\
Let a two-dimensional metric tensor field, which in a local coordinate system
$\{x^{\mu}\}\equiv\{u,v\}$, has the form:
\be
g_{\alpha\beta}(u,v;\lambda)=(1+\lambda^{2})
\begin{pmatrix}
 0 & 1+u^{2}+(1+\lambda^{2})^{2}v^{2} \\
 1+u^{2}+(1+\lambda^{2})^{2}v^{2} & 0
\end{pmatrix}
\ee
Solution to:
\be
N_{\alpha|\beta}+N_{\beta|\alpha}=\dot{g}_{\alpha\beta}
\ee
results in:
\be
N^{\alpha}=\Big\{0,\frac{2\lambda v}{1+\lambda^{2}}\Big\}\equiv
\Big\{0,\frac{2y^{0} y^{2}}{1+(y^{0})^{2}}\Big\}
\ee
and hence:
\be
n^{A}=\frac{1}{N(y^{0})}\Big\{1,0,-\frac{2y^{0} y^{2}}{1+(y^{0})^{2}}\Big\}
\ee
The corresponding flow lines are described by:
\bse
\begin{align}
&\frac{dy^{0}}{d s}=\frac{1}{N(y^{0})}\\
&\frac{dy^{1}}{d s}=0\\
&\frac{dy^{2}}{d s}=-\frac{1}{N(y^{0})}\frac{2y^{0}y^{2}}{(1+(y^{0})^{2})}
\end{align}
\ese
and the integral curves:
\bse
\begin{align}
&\int N(y^{0})dy^{0}=s+\overline{y}^{0}\\
&y^{1}=\overline{y}^{1}\\
&y^{2}=\overline{y}^{2}(1+(y^{0})^{2})^{-1}
\end{align}
\ese
Then, as expected, it is:
\be
\overline{n}^{A}=\Big\{1,0,0\Big\}
\ee
leading to the transformed embedding metric:
\be
\overline{\widetilde{g}}_{AB}=\begin{pmatrix}
  \varepsilon & 0 \\
  0 & \overline{g}_{\alpha\beta}
\end{pmatrix}
\ee
with:
\be
\overline{g}_{\alpha\beta}=\begin{pmatrix}
0 & 1+\overline{u}^{2}+\overline{v}^{2} \\
 1+\overline{u}^{2}+\overline{v}^{2} & 0
\end{pmatrix}
\ee

Though the example may seem simple and trivial, its purpose is to exhibit not only the
implementation of the criterion but also all the details connected to it.

\vspace{1cm}
\noindent
\textbf{Acknowledgements}

The project is co-funded by the European Social Fund and National Resources ---(EPEAEK II)---
PYTHAGORAS II.\\
The author wishes to thank Associate Professor Dr.\ T. Chistodoulakis, Dr.\ S. Bonanos, and
M.Sc.\ P. Terzis for stimulating discussions.

%%%%%%%%%%%%%%%%%%%%%%%%%%%%%%%%%%%%% Bibliography %%%%%%%%%%%%%%%%%%%%%%%%%%%%%%%%%%%%%%%%%%%%%%

\end{document}